# Moiré quasi-bound states in the continuum

Lei Huang[*], Weixuan Zhang[*], and Xiangdong Zhang[$]

Key Laboratory of advanced optoelectronic quantum architecture and measurements of Ministry of Education, Beijing Key Laboratory of Nanophotonics & Ultrafine Optoelectronic Systems, School of Physics, Beijing Institute of Technology, 100081, Beijing, China

[*]These authors contributed equally to this work. [$]Author to whom any correspondence should be addressed. E-mail: zhangxd@bit.edu.cn

## Abstract

The novel physics of twisted bilayer graphene has motivated extensive studies of magic-angle flat bands hosted by moiré structures in electronic, photonic and acoustic systems. On the other hand, bound states in the continuum (BICs) have also attracted great attention in recent years because of their potential applications in the field of designing superior optical devices. Here, we combine these two independent concepts to construct a new optical state in a twisted bilayer photonic crystal slab, which is called as moiré quasi-BIC, and numerically demonstrate that such an exotic optical state possesses dual characteristics of moiré flat bands and quasi-BICs. To illustrate the mechanism for the formation of moiré flat bands, we develop an effective model at the center of the Brillouin zone and show that moiré flat bands could be fulfilled by balancing the interlayer coupling strength and the twist angle around the band edge above the light line. Moreover, by decreasing the twist angle of moiré photonic crystal slabs with flat bands, it is shown that the moiré flat-band mode at the Brillouin center gradually approaches a perfect BIC, where the total radiation loss from all diffraction channels is significantly suppressed. To clarify the advantage of moiré quasi-BICs, enhanced second-harmonic generation (SHG) is numerically proven with a wide-angle optical source. The efficiency of SHG assisted by designed moiré quasi-BICs can be greatly improved compared with that based on dispersive quasi-BICs with similar quality factors.

Recently, there has been a great deal of interest in studying moiré superlattices formed in twisted bilayer van der Waals structures [1-20]. Analogous to moiré bilayers in condensed matter physics, the study of photonic moiré superlattices, which incorporate the twist degree of freedom into the photonic structure, has also received an increasing amount of attention [21-33]. For instance, the localization-delocalization transition of light has been demonstrated based on optical moiré superlattices [22, 23]. The topological transition of iso-frequency curve is observed in twisted bilayer α-$MoO_3$ flakes [24]. Additionally, there are also some direct analogs of twisted bilayer graphene using phononic and photonic crystal (PhCs) slabs [28-33] and bilayer electric circuits [34]. These twist-enabled wave phenomena could inspire us to design next-generation optical devices with novel performances.

On the other hand, bound states in the continuum (BICs), another fascinating research field in photonics, have also attracted great interest in recent years. A perfect optical BIC, which has an infinite quality factor (Q-factor), could exist in lossless photonic structures without radiations to the surrounding environment [35-55]. In practice, the applicable BIC-based optical modes should possess finite Q-factors and resonance widths to be excited. This could always be fulfilled by weakly deviating from perfect BICs, where the incident wave vector or the spatial symmetry is slightly changed. Recent investigations have shown that high-Q quasi-BICs can be realized in various nanostructures, and many applications, such as enhanced nonlinear effects [53] and low-threshold nanolasers [54,55], can be achieved. Although these fascinating applications have already been fulfilled, the dispersive effect should limit the performance of quasi-BICs under wide-angle illuminations, where resonant frequencies associated with the input signal along different propagation directions deviate from each other. The question is whether moiré physics and BICs can be combined to construct novel optical states, which are superior to current quasi-BICs and could further improve the optical efficiency in many applications.

In general, it is difficult to fulfill moiré quasi-BICs with flat-band dispersion in PhCs. Because, most investigations on engineering two-dimensional optical moiré flat bands directly refer to the electronic counterpart, where the Dirac point at the corner of the Brillouin zone (BZ) is necessary. However, the ideal Dirac point is hard to find above the light line due to the complex mode coupling in high-frequency ranges. In this case, current investigations on the design of moiré flat bands in PhC slabs are all in the region below the light line. Paradoxically, quasi-BICs in PhC slabs must locate above the light line. This makes it very difficult to combine BICs and moiré flat bands to construct moiré quasi-BICs. Consequently, a new method of designing moiré flat bands above the light line must be created.

In this work, we propose an effective model at the band edge of twisted bilayer PhC slabs to construct a moiré quasi-BIC above the light line. It is found that a moiré quasi-BIC with a narrow band dispersion could be achieved by engineering the interlayer coupling and twisting of bilayer PhC slabs. Moreover, we numerically calculate the band dispersion and find that the calculated energy bands are consistent with the effective model. In particular, by decreasing the twist angle, the moiré quasi-BIC in a flat band gradually approaches a perfect BIC. With a wide-angle optical source, we numerically demonstrate that the efficiency of SHG assisted by the moiré quasi-BIC can be greatly improved compared to that based on dispersive quasi-BICs with a similar Q-factor.

***Engineering moiré flat bands above the light line.*** — We start by considering the system composed of aligned bilayer PhC slabs suspended in air, as shown in the top chart of Fig. 1(a). The period and thickness of the PhC slab are chosen to be a=720 nm and d=144 nm. The radius of air holes is r=144 nm. The material of PhC slabs is GaAs with *n*=3.4 [56]. As shown in Fig. 1(b), we calculate the energy band of the bilayer structure with an extremely large interlayer distance (L>>a). In this case, the interlayer coupling is nearly negligible, making the energy band of bilayer PhC slabs become a twofold degeneration of the single-layer counterpart. It is worthwhile to note that each energy band has a quadratic dispersion with respect to the Γ point, which possesses fourfold degeneration, and red and blue lines are used to mark the low- and high-frequency energy bands, respectively.

Due to the orthogonality of low- and high-frequency eigenmodes, the interlayer coupling (*L~a*) of bilayer PhC slab could only exist between either a pair of low-frequency or high-frequency energy bands. Here, we focus on the construction of an effective model for low-frequency energy bands, which are crucial for the formation of moiré flat bands, as demonstrated below. In this case, the effective Hamiltonian of a pair of low-frequency bands is expressed as

$$H_{bilayer} = \begin{pmatrix} \omega_0 + bk^2 & U \\ U & \omega_0 + bk^2 \end{pmatrix}, \tag{1}$$

where $\omega_0$ is the eigenfrequency of the low-frequency band at the Γ point of a single-layer PhC slab. *b* is the dispersion coefficient and *U* is the interlayer coupling strength, which increases exponentially as the interlayer distance decreases. The detailed deviation of Eq. (1) is provided in S1 of Ref. [57].

Next, we rotate the bilayer PhC slab by an angle *θ* with respect to a common air hole to achieve moiré PhC slabs. To ensure the translational symmetry, the commensurate rotation angle should be defined as cos(*θ*) = ($m^2 + n^2 +$ 4*mn*)/($2m^2 + 2n^2 + 2mn$) with *m* and *n* being two integers. The bottom chart in Fig. 1a plots the moiré unit cell with

$\theta=9.43°$ ($m=3$ and $n=4$), and $a_1$ (and $a_2$) represents the real-space lattice vector. Fig. 1c shows the BZ of the moiré superlattice in the extended scheme. Two red and blue large hexagons correspond to the first BZs of the top and bottom layers, where K' and K'' denote two equivalent valleys of each layer. Black small hexagons correspond to the folded BZ of the moiré PhC slab, where $\Gamma^0$, $\Gamma^1$, $\Gamma^2$, and $\Gamma^3$ label center points in the first-, second-, third- and fourth-order moiré BZs. It is noted that low-frequency quadratic dispersions could also exist at each $\Gamma$ point of the higher-order moiré BZs.

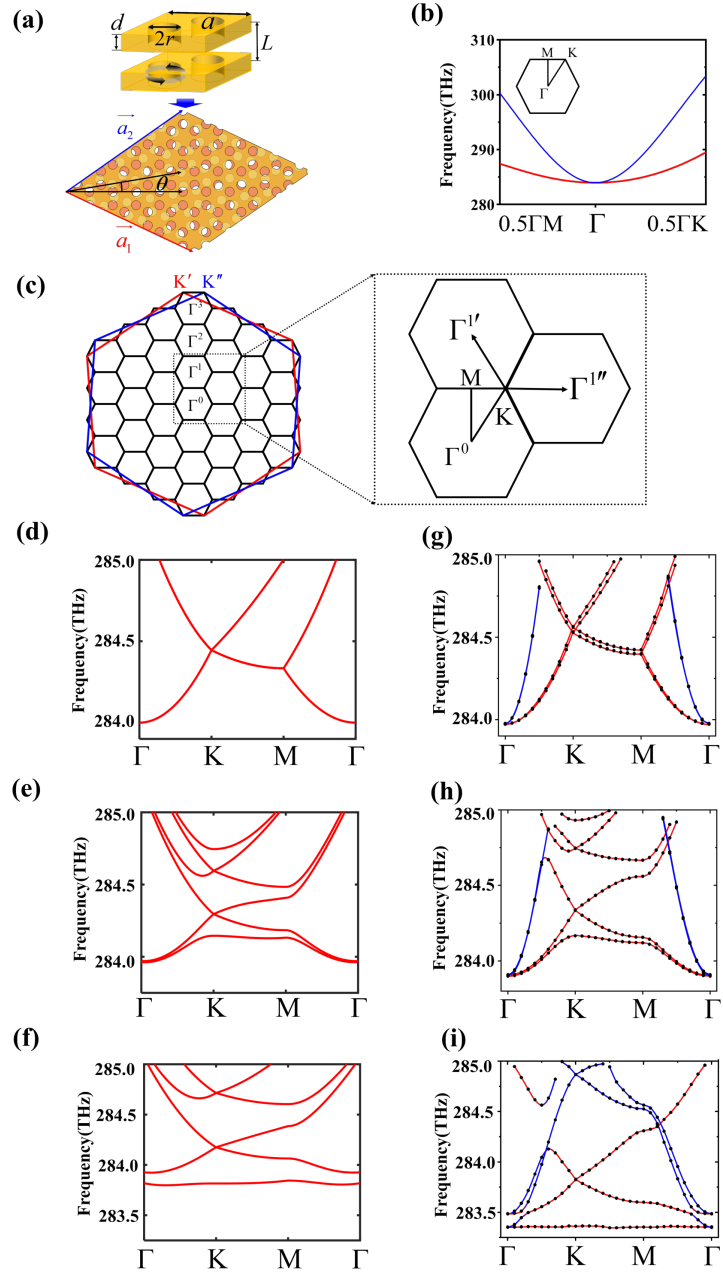

FIG. 1. (a) The top and bottom charts present the unit cell of aligned and twisted bilayer PhC slabs with $\theta=9.43°$. (b) Energy bands of the bilayer PhC slab with an large interlayer distance. (c) Schematic diagram of the moiré BZ in the extended scheme. (d-f) Band dispersions with three interlayer coupling strengths of the effective model. (g)-(i) Calculated

band structures of moiré PhC slabs with interlayer distances being 1500 nm, 600 nm and 470 nm. Here, the twist angle is $\theta=9.43°$. There is an anti-crossing induced by the coupling between the high-energy (blue line) and the low-energy (red line) bands.

To further illustrate the property of low-frequency moiré bands, we construct an effective model of moiré PhC slabs around the center of the first moiré BZ. To accurately describe the low-frequency band dispersion at the K point shown in the enlarged view of Fig. 1c, eigenmodes belonging to the three nearest moiré BZs should be considered in the effective model. This is because eigenmodes originating from other high-order moiré BZs around the K point possess much larger frequencies, making the influence of other BZs on the low-frequency moiré bands become negligible. In this case, the basis function of the effective mode can be given by $\psi_{moiré} = \left(E_1^{(1)} \quad E_1^{(2)} \quad E_2^{(1)} \quad E_2^{(2)} \quad E_3^{(1)} \quad E_3^{(2)}\right)^T$, where the subscript labels the low-frequency mode of three nearest moiré BZs, and the superscript represents the upper and lower PhC slabs. Based on this six-band basis, the low-frequency effective Hamiltonian of moiré PhC slabs can be written as

$$H_{moiré} = \begin{pmatrix} H_1 & V_0 & V_0 \\ V_0 & H_2 & V_0 \\ V_0 & V_0 & H_3 \end{pmatrix} \quad H_\xi = \begin{pmatrix} \omega_0 + b(\vec{k} - \vec{G_\xi})^2 & U \\ U & \omega_0 + b(\vec{k} - \vec{G_\xi})^2 \end{pmatrix} \quad V_0 = \begin{pmatrix} 0 & V \\ V & 0 \end{pmatrix}, \quad (2)$$

where $\xi = 1, 2, 3$ and distances from three nearest BZs to the center of k-space equal to $\vec{G_1} = (0,0)$, $\vec{G_2} = \overline{\Gamma^0 \Gamma^{1'}}$, and $\vec{G_3} = \overline{\Gamma^0 \Gamma^{1''}}$. In particular, three two-by-two diagonal matrices represent effective Hamiltonians of moiré bands in three nearest BZs, where $U$ is the interlayer coupling from the same BZ. The off-diagonal matrix $V_0$ corresponds to the interlayer coupling between moiré bands from different BZs ($V$ is the coupling strength). Such long-range interlayer couplings result from the twist-induced reduction of the original translation symmetry. It is noted that in the small twist-angle limitation, the value of interlayer coupling strength is independent of the *k* vector. Hence, we could treat the interlayer coupling strengths as constants.

Based on the effective model, in Figs. 1d-1f, we calculate band dispersions with three different interlayer couplings at $\theta=9.43°$. As plotted in Fig. 1d, when the interlayer coupling is zero, the band of moiré PhC slab is identical with the folded band of the single-layer PhC slab. With increasing interlayer couplings (U=0.003, V=0.03), the band degeneracy at K point is opened shown in Fig. 1e. This is because the required translational symmetry, which protects the degeneracy at K point, is destroyed by the non-ignored interlayer coupling. By further increasing the coupling strength to a magic value (U=0.01, V=0.1), the lowest band becomes very flat, as shown in Fig. 1f. In this case, we can see that the low-frequency moiré flat band could appear with the balance between the twist angle and the interlayer coupling in moiré PhC slabs. Specifically, unlike the effective model of twisted bilayer graphene around the BZ corner of the graphene monolayer, here, we formulate an effective model around the BZ center. It is widely known that all eigenmodes of the PhC slab at the BZ center locate above the light line. Hence, our effective model gives rise to a new method to engineer moiré flat bands above the light line. In addition, it is worthy to note that the relationship between different orders of Fourier components for the interlayer coupling strength ($U$ and $V$) is different from that of twisted bilayer graphene. This is due to the fact that the coupling between two PhC slabs with a relatively large distance (the coupling induced by the evanescent fields is weak) depends on the leaky rate of optical modes. See S2 of Ref. [57] for details.

To test the validity of the above proposed effective model, we numerically calculate the band structure of moiré PhC slabs with $\theta=9.43°$. The result with $L=1500$ nm is presented in Fig. 1g. Red and blue lines correspond to bands rooted from the low- and high-frequency bands of the aligned bilayer PhC slab. It is seen that the low-frequency part of band structure is similar to that predicted by the effective model without interlayer couplings (in Fig. 1d). The difference could be interpreted by the existence of very weak interlayer couplings in the real structure, which slightly breaks the degeneration at K point. By decreasing the interlayer distance to $L=600$ nm, the effective interlayer coupling becomes stronger. The calculated band dispersion is displayed in Fig. 1h, where two pairs of degenerated modes at K point appear in separated frequencies and the lowest band becomes flatter. These features could be accurately manifested by the effective model with suitable interlayer couplings (in Fig. 1e). When the interlayer distance is tuned to $L=470$ nm, the interlayer coupling strength further increases, making the lowest band become a nearly perfect moiré flat band, as shown in Fig. 1i. Comparing the calculated band with that given by the effective model, we obtain good consistency around the K point. The slight difference around M and Γ points is due to the approximation with three moiré BZs in the effective model. In S2 of Ref. [57], an effective model considering four nearest BZs is proposed, which can give more accurate results around M point.

In S3 of Ref. [57], band dispersions with the twist angle being 7.34° are calculated at different interlayer distances. The variation of band flatness as functions of the interlayer distances are also provided at three twist angles 7.43°, 9.43° and 13.2° [57]. Based on analytical and numerical results, we can see that moiré flat bands could be fulfilled by balancing the interlayer coupling strength and the twist angle. Specifically, the smaller the twist angle is, the lower the interlayer coupling strength is required to form the flat-band moiré BIC. In this case, when the twist angle is fixed, there is a suitable interlayer coupling strength (interlayer distance) to balance the twist angle to form moiré flat bands. Referring to above results, it is expected that our proposed effective model provides a guideline to engineer moiré flat bands above the light line, and in the following, we show that such a method could be used to achieve the combination of moiré flat bands and BICs.

*Moiré quasi-BICs in twisted bilayer PhC slabs.*—Previous investigations have shown that symmetry-protected BICs always exist at the BZ center. For the above aligned bilayer PhC slab, the $C_6$ symmetry at the $\Gamma$ point is expected to induce a twofold-degenerate BIC. In Fig. 2a, we calculate the far-field polarization states and Q-factors of the bilayer PhC slab, where white lines represent the linear polarization and the color bar measures Q-factors. It is shown that the far-field polarization state at the $\Gamma$ point is ill-defined, making the radiation field vanish and the topological charge equal to -2. In addition, the Q-factor is exponentially increased to infinite when the *k*-vector approaches to $\Gamma$ point. In this case, the mode at the BZ center is a symmetry-protected BIC.

By twisting the bilayer PhC slab, an enlarged moiré supercell could be realized. It is noted that the period of moiré unit cell is much larger than the wavelength. In this case, except for the zero-order diffraction channel, other higher-order diffraction channels appear. Figs. 2b and 2c illustrate schematic diagrams of zero-order and first-order diffraction channels with $\theta$=9.43° and L=470 nm. To analyze the far-field radiation property, we calculate the zero-order and first-order far-field polarization states toward the upward direction, as shown in Figs. 2d and 2e. Results for far-field polarization states toward the downward direction are shown in S5 of Ref. [57]. Red and blue ellipses denote the right- and left-handed elliptic polarizations. We can see that due to the twist-induced breaking of updown symmetry, zero-order far-field polarization states are changed from linear polarizations to elliptic polarizations with different major axis orientations. In addition, it is clearly shown that the far-field polarization state at $\Gamma$ point of the zero-order diffraction channel is still ill-defined. For the six first-order diffraction channels, there are certain polarization states at $\Gamma$ point, indicating that eigenmodes at the BZ center could couple with the environment. It is worth noting that, similar to the first-order channels, other existing higher-order channels could also couple with radiational modes.

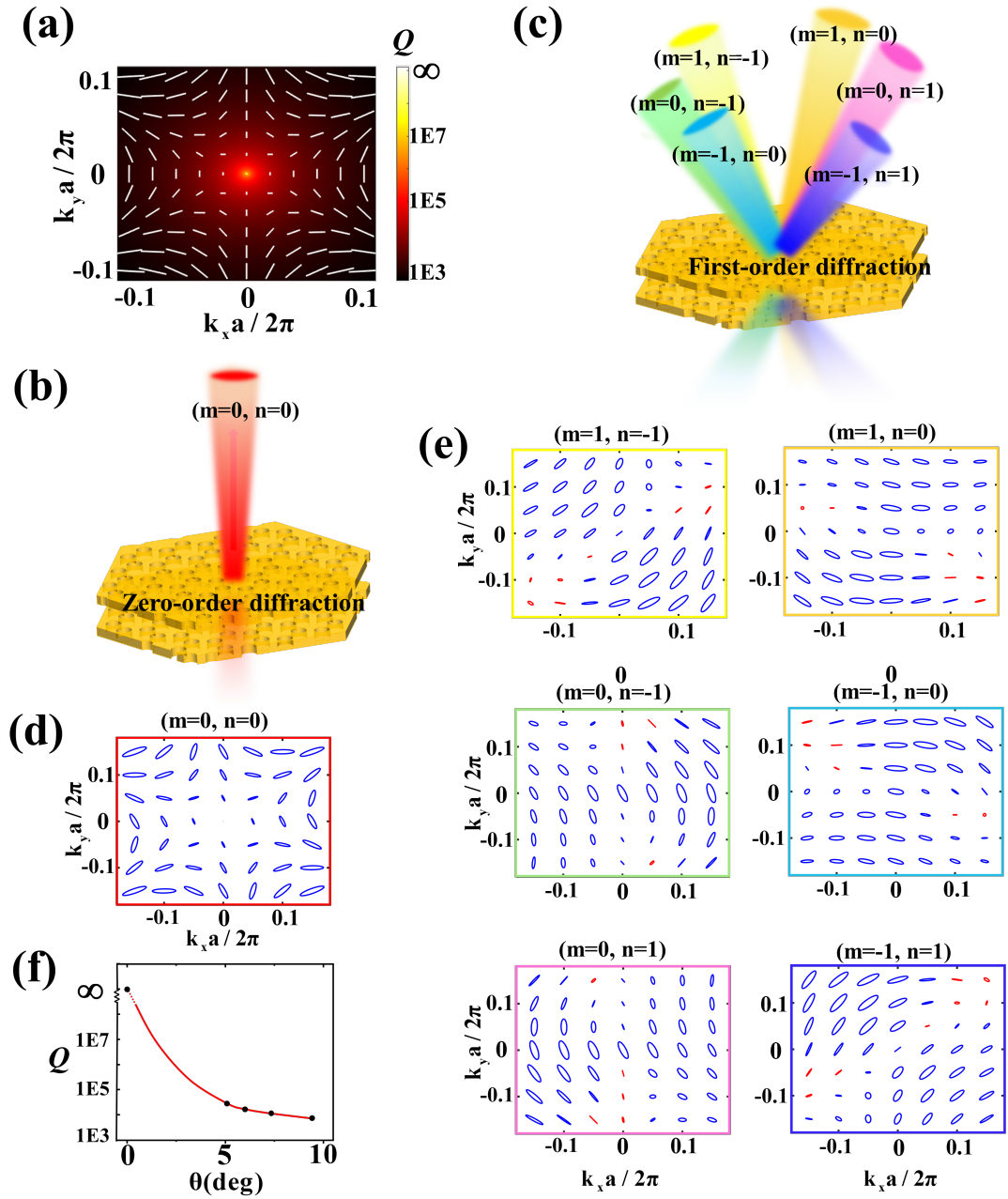

FIG. 2. (a). Far-field polarization states and Q-factors of the aligned bilayer PhC slab. (b) and (c). Schematic diagrams of zero-order and first-order diffraction channels with $\theta=9.43°$. (d) and (e). Far-field polarization states of zero-order and first-order diffraction channels toward the upward direction with with $\theta=9.43°$ and L=470 nm. (f). Q-factors of moiré quasi-BICs as a function of the twist angle. The interlayer distance is different for each twist angle to fulfill the flat band.

Although these higher-order diffraction channels could make the original perfect BIC in aligned bilayer PhC slabs become lossy, we find that by decreasing the twist angle, the Q-factor at BZ center for the low-frequency flat band is exponentially increased to infinite, as shown in Fig. 2f. This could be explained by the much weaker interlayer coupling for the realization of low-frequency moiré flat bands when the twist angle decreases. The weak interlayer coupling could decrease the radiational loss from higher-order diffraction channels. In this case, the eigenmode at the BZ center of the moiré flat band evolves from a perfect BIC. Hence, we call these modes as moiré quasi-BICs, which is beneficial for various applications under a wide-angle illumination. As an instance, we numerically prove that the nonlinear effect could be greatly enhanced compared to that based on dispersive quasi-BICs with similar Q-factors.

***Enhanced second-harmonic generation by moiré quasi-BICs.***—Next, we provide a numerical demonstration of the superior performance in the enhancement of SHG by flat-band moiré quasi-BICs. Here, the incident light is regarded as a wide-angle source, as illustrated in Fig. 3a. The angular-resolved intensity is written as $I(\theta_{in}) = I_0 \exp(-\gamma \theta_{in}^2)$, where $\theta_{in}$ marks the incident angle and $\gamma = 32$ determines the angular spectrum. For comparison, we first consider the moiré PhC slab with $\theta=13.2°$ and L=460 nm, which possesses a dispersive quasi-BIC with a Q-factor similar to designed moiré quasi-BIC with flat bands. As shown in Fig. 3b, we calculate the average efficiency of SHG, which is defined as $\eta_{ave} = \iint I(\theta_{in})\eta(\theta_{in},\varphi)d\theta_{in}d\varphi / \iint I(\theta_{in})d\theta_{in}d\varphi$ with $\eta(\theta_{in},\varphi)$ being the efficiency of SHG with a fixed incident angle $(\theta_{in},\varphi)$. The inset displays the efficiency of SHG at different incident angles. We can see that due to the dispersion effect, the maximum values of $\eta(\theta_{in},\varphi)$ at different incident angles deviate from each other, extending the total nonlinear signal over a wide frequency range.

Then, we calculate the average efficiency of SHG using the above designed twisted bilayer PhC slab sustaining moiré quasi-BIC with a flat band ($\theta=9.43°$ and L=470 nm) shown in Fig. 3c. The inset shows the efficiency of SHG at different incident angles. We can see that the maximum values of $\eta(\theta_{in},\varphi)$ at different incident angles merge at a single frequency. In this case, the total nonlinear signal at the flat-band frequency is further enlarged, which is nearly ten times larger than that based on dispersive quasi-BICs. Hence, we can deduce that moiré quasi-BICs could exhibit highly efficient performances in various surface-enhanced applications. In addition, by considering the suitable Q-factor, which is high enough for applications and also possesses the excitable resonance width, and the size of twisted bilayer PhCs with a sufficient amount of moiré units, we suggest the twist angle could be set in the range from 4.41° to 9.43°.

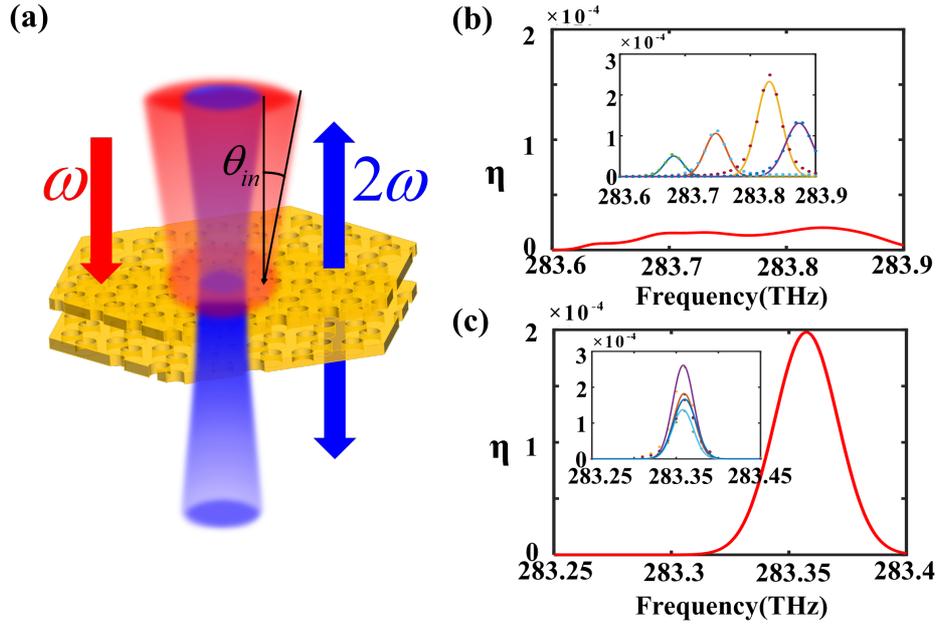

FIG. 3. (a). Scheme of enhanced SHG by moiré quasi-BICs under a wide-angle source. The average efficiency of SHG based on dispersive ($\theta$=13.2° and L=460 nm) moiré quasi-BIC for (b), and flat-band ($\theta$=9.43° and L=470 nm) moiré quasi-BIC for (c). Insets present efficiencies of SHG with the incident angles being $\theta_{in} = 2.86°, 5.73°, 8.59°$ and $11.46°$.

In conclusion, we have proposed an effective model of twisted bilayer PhC slabs to construct a moiré quasi-BIC around the center of the BZ. We find that a moiré quasi-BIC with a narrow band dispersion could be achieved by engineering the interlayer coupling and twisting of the bilayer PhC slabs. Moreover, the calculated bands are consistent with those obtained from the effective model. By decreasing the twist angle of the bilayer PhC slab with a flat band, the moiré quasi-BIC gradually approaches a perfect BIC. Furthermore, with a wide-angle optical source, we demonstrate that the efficiency of SHG assisted by moiré quasi-BICs can be greatly enhanced compared to that based on dispersive quasi-BICs with similar Q-factors. Our findings may have potential applications in designing BIC-based photonic devices with high efficiencies.

This work was supported by the National key R & D Program of China under Grant No. 2017YFA0303800 and the National Natural Science Foundation of China (No.91850205).

[1] J. M. Lopes Dos Santos, N. M. Peres, and A. H. Castro Neto, Phys. Rev. Lett. **99**, 256802 (2007).
[2] G. Trambly de Laissardiere, D. Mayou, and L. Magaud, Nano Lett **10**, 804 (2010).
[3] E. Suárez Morell, J. D. Correa, P. Vargas, M. Pacheco, and Z. Barticevic, Phys. Rev. B **82**, 121407(R) (2010).

[4] R. Bistritzer and A. H. MacDonald, Proc Natl Acad Sci U S A **108**, 12233 (2011).
[5] Y. Cao, V. Fatemi, S. Fang, K. Watanabe, T. Taniguchi, E. Kaxiras, and P. Jarillo-Herrero, Nature **556**, 43 (2018).
[6] Y. Cao *et al.*, Nature **556**, 80 (2018).
[7] X. Lu *et al.*, Nature **574**, 653 (2019).
[8] Y. Jiang, X. Lai, K. Watanabe, T. Taniguchi, K. Haule, J. Mao, and E. Y. Andrei, Nature **573**, 91 (2019).
[9] K. Tran *et al.*, Nature **567**, 71 (2019).
[10] Y. Choi *et al.*, Nature **589**, 536 (2021).
[11] A. Kerelsky *et al.*, Nature **572**, 95 (2019).
[12] C. Jin *et al.*, Nature **567**, 76 (2019).
[13] E. M. Alexeev *et al.*, Nature **567**, 81 (2019).
[14] K. L. Seyler, P. Rivera, H. Yu, N. P. Wilson, E. L. Ray, D. G. Mandrus, J. Yan, W. Yao, and X. Xu, Nature **567**, 66 (2019).
[15] M. Yankowitz, S. Chen, H. Polshyn, Y. Zhang, K. Watanabe, T. Taniguchi, D. Graf, A. F. Young, and C. R. Dean, Science **363**, 1059 (2019).
[16] A. Uri *et al.*, Nature **581**, 47 (2020).
[17] U. Zondiner *et al.*, Nature **582**, 203 (2020).
[18] Y. Tang *et al.*, Nature **579**, 353 (2020).
[19] G. Chen *et al.*, Nature **579**, 56 (2020).
[20] X. Liu *et al.*, Nature **583**, 221 (2020).
[21] K. Dong *et al.*, Phys. Rev. Lett. **126**, 223601 (2021).
[22] P. Wang, Y. Zheng, X. Chen, C. Huang, Y. V. Kartashov, L. Torner, V. V. Konotop, and F. Ye, Nature **577**, 42 (2020).
[23] Q. Fu, P. Wang, C. Huang, Y. V. Kartashov, L. Torner, V. V. Konotop, and F. Ye, Nature Photonics **14**, 663 (2020).
[24] G. Hu *et al.*, Nature **582**, 209 (2020).
[25] S. Sunku *et al.*, Science **362**, 1153 (2018).
[26] D. X. Nguyen, X. Letartre, E. Drouard, P. Viktorovitch, H. C. Nguyen, and H. S. Nguyen, arXiv preprint arXiv:2104.12774 (2021).
[27] B. Lou, N. Zhao, M. Minkov, C. Guo, M. Orenstein, and S. Fan, Phys. Rev. Lett. **126**, 136101 (2021).
[28] J. Lu, C. Qiu, W. Deng, X. Huang, F. Li, F. Zhang, S. Chen, and Z. Liu, Phys. Rev. Lett. **120**, 116802 (2018).
[29] Y. Deng, M. Oudich, N. J. R. K. Gerard, J. Ji, M. Lu, and Y. Jing, Phys. Rev. B **102**, 180304(R) (2020).
[30] M. Rosendo Lopez, F. Penaranda, J. Christensen, and P. San-Jose, Phys. Rev. Lett. **125**, 214301 (2020).
[31] H. Tang, F. Du, S. Carr, C. DeVault, O. Mello, and E. Mazur, Light Sci Appl **10**, 157 (2021).
[32] M. Oudich, G. Su, Y. Deng, W. Benalcazar, R. Huang, N. J. Gerard, M. Lu, P. Zhan, and Y. Jing, arXiv preprint arXiv:2103.03686 (2021).
[33] C.-H. Yi, H. C. Park, and M. J. Park, arXiv preprint arXiv:2108.14002 (2021).
[34] W. Zhang, D. Zou, Q. Pei, W. He, H. Sun, and X. Zhang, Phys. Rev. B **104**, L201408 (2021).
[35] C. W. Hsu, B. Zhen, J. Lee, S. L. Chua, S. G. Johnson, J. D. Joannopoulos, and M. Soljacic, Nature **499**, 188 (2013).
[36] C. W. Hsu, B. Zhen, A. D. Stone, J. D. Joannopoulos, and M. Soljačić, Nature Reviews Materials **1** (2016).
[37] B. Zhen, C. W. Hsu, L. Lu, A. D. Stone, and M. Soljacic, Phys. Rev. Lett. **113**, 257401 (2014).
[38] Y. Yang, C. Peng, Y. Liang, Z. Li, and S. Noda, Phys. Rev. Lett. **113**, 037401 (2014).
[39] J. Jin, X. Yin, L. Ni, M. Soljacic, B. Zhen, and C. Peng, Nature **574**, 501 (2019).

[40] H. M. Doeleman, F. Monticone, W. den Hollander, A. Alù, and A. F. Koenderink, Nature Photonics **12**, 397 (2018).
[41] X. Yin, J. Jin, M. Soljacic, C. Peng, and B. Zhen, Nature **580**, 467 (2020).
[42] B. Wang *et al.*, Nature Photonics **14**, 623 (2020).
[43] A. Regensburger, M. A. Miri, C. Bersch, J. Nager, G. Onishchukov, D. N. Christodoulides, and U. Peschel, Phys. Rev. Lett. **110**, 223902 (2013).
[44] S. I. Azzam, V. M. Shalaev, A. Boltasseva, and A. V. Kildishev, Phys. Rev. Lett. **121**, 253901 (2018).
[45] Y. Zhang *et al.*, Phys. Rev. Lett. **120**, 186103 (2018).
[46] M. Kang, S. Zhang, M. Xiao, and H. Xu, Phys. Rev. Lett. **126**, 117402 (2021).
[47] W. Liu, B. Wang, Y. Zhang, J. Wang, M. Zhao, F. Guan, X. Liu, L. Shi, and J. Zi, Phys. Rev. Lett. **123**, 116104 (2019).
[48] S. Vaidya, W. A. Benalcazar, A. Cerjan, and M. C. Rechtsman, Phys. Rev. Lett. **127**, 023605 (2021).
[49] Y. Plotnik, O. Peleg, F. Dreisow, M. Heinrich, S. Nolte, A. Szameit, and M. Segev, Phys. Rev. Lett. **107**, 183901 (2011).
[50] S. D. Krasikov, A. A. Bogdanov, and I. V. Iorsh, Journal of Physics: Conference Series **1092**, 012068 (2018).
[51] T. Yoda and M. Notomi, Phys. Rev. Lett. **125**, 053902 (2020).
[52] A. Kodigala, T. Lepetit, Q. Gu, B. Bahari, Y. Fainman, and B. Kante, Nature **541**, 196 (2017).
[53] S. Joseph, S. Pandey, S. Sarkar, and J. Joseph, Nanophotonics **10**, 4175 (2021).
[54] C. Huang *et al.*, Science **367**, 1018 (2020).
[55] K. Hirose, Y. Liang, Y. Kurosaka, A. Watanabe, T. Sugiyama, and S. Noda, Nature Photonics **8**, 406 (2014).
[56] W. Zhang *et al.*, Light Sci. Appl. **9**, 109 (2020).
[57] See Supplemental Material for (S1) the effective model of bilayer photonic crystal slab, (S2) the effective model of moiré photonic crystal slab, (S3) the energy band of the moiré photonic crystal slab, (S4) far-field polarization states toward the downward direction of the moiré photonic crystal slab.